\documentclass[debug]{rmaa}

\usepackage{paralist}

\usepackage{psfrag,color}

\usepackage[utf8]{inputenc}

\usepackage{multirow}

\title{Parameters of Recent Transits of HAT-P-23\lowercase{b}} 

\author{
  F. G. Ramón-Fox,\altaffilmark{1}
  P. V. Sada,\altaffilmark{2}}

\altaffiltext{1}{Tecnológico de Monterrey, México.}
\altaffiltext{2}{Universidad de Monterrey, México.}

\shortauthor{Ramón-Fox \& Sada}
\shorttitle{Parameters of Recent Transits of HAT-P-23\lowercase{b}}

\fulladdresses{
\item Pedro Valdés Sada: Universidad de Monterrey, 66238, San Pedro Garza García, Nuevo León, México (pedro.valdes@udem.edu.mx).
\item Felipe G. Ramón Fox: Instituto Tecnológico y de Estudios Superiores de Monterrey, 64849, Monterrey, Nuevo León, México. (fg.ramon.fox@googlemail.com).}

\listofauthors{F. G. Ramón-Fox \& P. V. Sada}
\indexauthor{Ramón-Fox, F. G.}
\indexauthor{Sada, P. V.}

\abstract{Four transits of the exoplanet HAT-P-23b were recently observed with the 0.36 m telescope at the Universidad de Monterrey Observatory. The four light curves were successfully combined to obtain a resulting one with reduced scattering per bin. This one was modeled using a Monte Carlo method to obtain the essential parameters that characterize the system. Assuming orbital parameters such as eccentricity $e$ and longitude of periastron $\omega$ from the discovery paper, we found values of $Rp/R_{\star} = 0.1105 ^{+0.0015}_{-0.0013}$ for the planet-to-star radius ratio, $a/R_{\star} = 4.23 ^{+0.06}_{-0.12}$ for the scaled semimajor axis, and an orbital inclination of the system of $i =  87.9 \arcdeg$~$ ^{+1.5}_{-2.2}$. We also derive an improved orbital period of $1.2128868 \pm 0.0000004$ days ($T_{\mathrm{o}} = 2,454,852.26542 \pm 0.00018$ BJD\_TDB) for the system. }

\resumen{Cuatro tránsitos del planeta extrasolar HAT-P-23b fueron observados recientemente con el telescopio de 0.36 m del Observatorio de la Universidad de Monterrey. Las cuatro curvas de luz fueron exitosamente combinadas para obtener una resultante de mayor calidad. Esta fue modelada utilizando un método de Monte Carlo para obtener los parámetros esenciales que caracterizan al sistema. Asumiendo parámetros orbitales como la excentricidad $e$ y la longitud del periastro $\omega$ reportados en el artículo del descubrimiento, encontramos valores de $Rp/R_{\star} = 0.1105 ^{+0.0015}_{-0.0013}$ para la razón del radio del planeta al radio estelar, $a/R_{\star} = 4.23 ^{+0.06}_{-0.12}$ para el semieje mayor normalizado, y una inclinación orbital de $i =  87.9 \arcdeg$~$ ^{+1.5}_{-2.2}$. Obtenemos un período orbital de $1.2128868 \pm 0.0000004$ días ($T_{\mathrm{o}} = 2,454,852.26542 \pm 0.00018$ BJD\_TDB).}

\addkeyword{Extrasolar Planets}
\addkeyword{Stars: Individual (HAT-P-23)}

\begin{document}
\maketitle

\section{Introduction}
\label{sec:intro}

HAT-P-23b is a relatively massive $(\sim 2 M_{\mathrm{JUP}})$ transiting exoplanet orbiting the G0 dwarf star GSC 1632-01396 $(m_{\mathrm{v}} \sim 12.4)$ on a close circular orbit with a period of $1.212884 \pm 0.000002$ days \citep{Bakos10}. In the discovery paper, the authors conclude that this exoplanet is an inflated hot Jupiter, which has one of the shortest characteristic in-fall times ($\sim 7.5 ^{+2.9}_{-1.8}$ Myr) before being engulfed by the star. The Rossiter-McLaughlin effect for HAT-P-23b was expected to be significant because of the moderately high rotational velocity of the star ($8.1\pm 0.5$ $\mathrm{km~s}^{-1}$) and transit depth ($\sim 17$ $\mathrm{mmag}$). However, \citet{Moutou11} measured a similar $v \sin i_s$ ($7.8 \pm 1.6$  $\mathrm{km~s}^{-1}$), but estimated a prograde aligned orbit with a projected angle between the orbital plane and the stellar equatorial plane of only $\lambda = +15\arcdeg \pm 22\arcdeg$ and suggested that the uncertainty in the inclination of the system has some impact on the relatively large error bars. Thus it is of importance to attempt to improve this parameter from light curve observations. It is also important to verify the radius of the planet since the current reported radius $R_p = 1.368 \pm  0.090 R_j$ \citep{Bakos10} cannot be reproduced by the theoretical models developed by \citet{Fortney08}, which suggest a smaller size for a planet of this mass. For planets that are larger than expected, models generally consider an internal source of heating to explain the increase in planetary radii \citep{Laughlin11}. Possible inflation mechanisms are tidal dissipation in orbit circularization \citep{Bodenheimer01}, kinetic energy transport from wind-driven activity in the atmosphere to dissipation at the interior \citep{GuillotShowman02}, and internal heating by Ohmic dissipation of currents driven to the interior, produced by magnetohydrodynamic mechanisms on the surface \citep{Batygin10, Laughlin11}.

In this paper, we analyze recent transits of the extrasolar planet HAT-P-23b in an attempt to improve the essential model parameters that characterize this system. We describe a method for reducing and combining the observations of each transit to obtain an average light curve from which the transit parameters can be obtained. In \S~\ref{sec:obs-and-reduction}, we discuss the methodology for the photometric observations and data reduction of the transits. In \S~\ref{sec:light-curve-analysis}, we describe our method for combining several transit curves in order to obtain a curve with reduced noise per bin, and we derive an improved orbital period for the system. In \S~\ref{sec:transit-modeling}, we obtain the transit parameters from a best-fit model. In \S~\ref{sec:discussion}, we compare our results with previous observations and in \S~\ref{sec:conclusion} we summarize our findings.

\section{Observations and Data Reduction}
\label{sec:obs-and-reduction}

We observed four transits of HAT-P-23b using the Universidad de Monterrey (UDEM) Observatory telescope on UT dates 2011 June 3, and on August 4, 16 and 21. This is a small private college observatory having Minor Planet Center Code 720 and is located at an altitude of 689 m, in the suburbs of Monterrey, México. The data were acquired using a standard Rc-band filter (630 nm) on a 0.36 m reflector with a $1280 \times 1024$ pixel CCD camera at 1.0 arcsec pixel$^{-1}$ scale, resulting in a field-of-view of $\sim 21.3 \times 17.1 \arcmin$. The observations were slightly defocused. This technique spreads light over more pixels, which allows longer exposure times without saturation and reduces systematic errors of focusing light on a few pixels, thus increasing photometric precision. Also, on-axis guiding was used to maintain pointing stability. The exposure times were $60$ s and the images were binned $2 \times 2$ to facilitate rapid readout $(\sim 3$ $ \mathrm{s})$. Each observing session lasted at least four hours in order to accommodate the transit event and also to cover about one hour before ingress and one hour after egress. Weather conditions were similar during all the observing sessions (clear and calm), with the August 4th night being a bit more windy and turbulent. In order to obtain accurate time stamps for our images, the data-taking computer was synchronized with Coordinated Universal Time (UTC) at the start of each observing session through the Internet. At the end of the observing session, time was independently verified, to the nearest second, using WWV radio time signals.

Standard dark current subtraction and twilight sky flat-field division process were performed on each image for calibration.  Aperture photometry was carried out on the target star and five comparison stars of similar $(\pm \sim 1.5)$ magnitude. The apertures used varied for each date due to the defocus and weather conditions, but were optimized to minimize the scatter of the resulting light curves. We found that slightly smaller scatter in the final light curves was obtained by averaging the magnitude differences of HAT-P-23 to each comparison star individually. This produced smaller scatter than the method of ratioing the target star flux to the sum of the fluxes of all comparison stars. We found that both methods were essentially equivalent when using a limited number of comparison stars with similar magnitudes. Our chosen method has the advantage of easily identifying and correcting unsuitable comparison star measurements (due to saturated pixel effects, faintness, cosmit rays, etc.) and also provides an independent measure of the uncertainty for each point, which was similar to the overall data scatter. We estimated the formal error for each HAT-P-23 photometric point as the standard deviation of the magnitude differences to the individual comparison stars, divided by the square root of their number (error of the mean).

After normalizing the target star to the comparison stars and averaging, some gradual variations as a function of time were found. This is perhaps caused by differential extinction between the transit and comparison stars, which generally have different and, in some cases, unknown spectral types. Bluer comparison stars are more affected by this atmospheric scattering effect than redder stars. Consequently, this variation was removed by using a linear air mass-dependent function of the form:
\begin{equation}
	\delta m = c(1 - X) + b
	\label{eq:amfit}				
\end{equation}
where $\delta m$ is the magnitude change applied to an individual measurement at air mass $X$, and $c$ and $b$ are best-fit constants necessary to remove the systematic effect and obtain a flat (horizontal) line for the out-of-transit baseline portions of the light curve. The best-fit model in equation (\ref{eq:amfit}) is determined from the out-of-transit data. All transits have at least one hour’s worth of baseline observations before the transit ingress and after the transit egress for this purpose. This could also have been corrected by introducing an intrinsic color term for the stars (using the 2MASS catalogue for example), but our chosen empirical procedure yields similar good results and is simpler to implement.

\begin{figure}[!h]
	\centering
	\includegraphics[width = 1.00\linewidth]{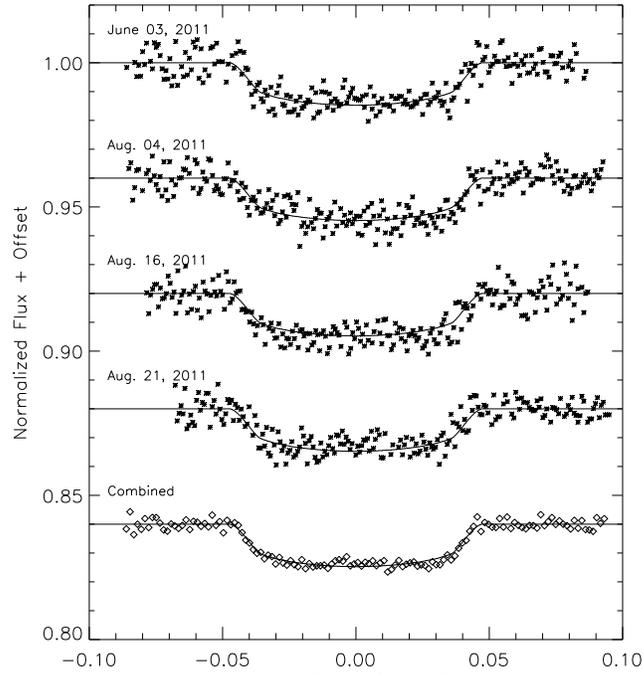}
	\caption{Individual light curves (top four with asterisks) observed through an Rc-band filter at UDEM Observatory, and final combined linghtcurve (bottom one with diamonds) with 2-minute bins. Superimposed is the best-fit model light curve used to determine the mid-transit times.}
	\label{fig:transits}
\end{figure}

The resulting light curves for the four observing dates are presented in Fig.~\ref{fig:transits}.

\section{light curve analysis}
\label{sec:light-curve-analysis}

\subsection{Light Curve Combination}

Transit photometry may present significant scattering due to various noise sources, specially in ground observations with small telescopes. Therefore, we propose to combine all the individual transit curves to produce a final average light curve with reduced noise per bin. This is warranted because all four light curves are from independent events, have similar scatter, the same exposure times, same number of comparison stars, and are devoid of star-spot activity. Combining all four transit light curves into one, should theoretically decrease the scatter by a factor of $\sim \sqrt{n}$. For our observations, this means a decrease by a factor of $\approx 2$. This procedure also has the tendency to cancel out any systematic effects that may be present in the individual light curves. From a computational point of view, this scheme facilitates the analysis of a relatively large number of subsequent follow-up photometry data, as the parameter space would increase by another mid-transit $T_{\mathrm{c}}$, making the model analysis more costly. 

In order to combine the light curves, we first need to co-register them by finding the mid-transit time for each event. In order to fit the observed transit light curves we first created initial standard model light curves. These were constructed numerically as a tile-the-star procedure using the Binary Maker II software \citep{Bradstreet05}. The initial system parameters used were from \citet{Bakos10} and the linear limb-darkening function coefficient ($u = 0.610$ for our bandpass) for this system was taken from \citet{Claret00}. Following a method similar to \citet{Sada12} and \citet{Todorov12}, small adjustments to the duration and depth of the model transits were necessary to optimize the fits and extract the best mid-transit time possible. This was done by applying small (a few percent) multiplicative factors to both the depth and duration of the model transit. Best-fit models were obtained by minimizing the $\chi^2$ of the data.  We used the scatter of the photometry to estimate the uncertainties because the scatter is larger than the formal errors suggest.

Once the initial mid-transit times for each system were obtained, a combined light curve was created by placing all the photometry points in a common time reference frame, centered on the mid-transit time for each event, and by averaging the data points falling within predetermined bins. This method generates a light curve with reduced noise per bin, equally spaced data points, and maintains the general shape of the transit curve. However, care must be taken in selecting the bin size in order to simultaneously have sufficient amount of data in each bin, and also sufficient data points to define the critical ingress and egress portions of the light curve. After testing different bin sizes, it was determined that 2-minute bins satisfied both criteria and this bin size was adopted for averaging the individual transits HAT-P-23b.

Further refinements were performed to the individual light curve mid-transit times by subjecting the combined light curve to the same $\chi^2$ fit to the preliminary model and obtaining general multiplicative factors to both the depth and duration of the model transit. These factors were then used on the individual light curves to derive improved mid-transit times and generate a new combined light curve. The process was iterated until there was convergence between the individual mid-transit times and the best-fit multiplicative factors for the combined light curve. Table~\ref{tab:mid-transit-times} presents the mid-transit times derived from fitting the models, as explained above, in the four observed transits for HAT-P-23, and Fig.~\ref{fig:transits} (bottom with diamond symbols) shows the resulting combined light curve for HAT-P-23.

\subsection{Ephemeris Determination}
\label{subsec:ephemeris}

\begin{figure}[!h]
	\centering
	\includegraphics[width = 0.95\linewidth]{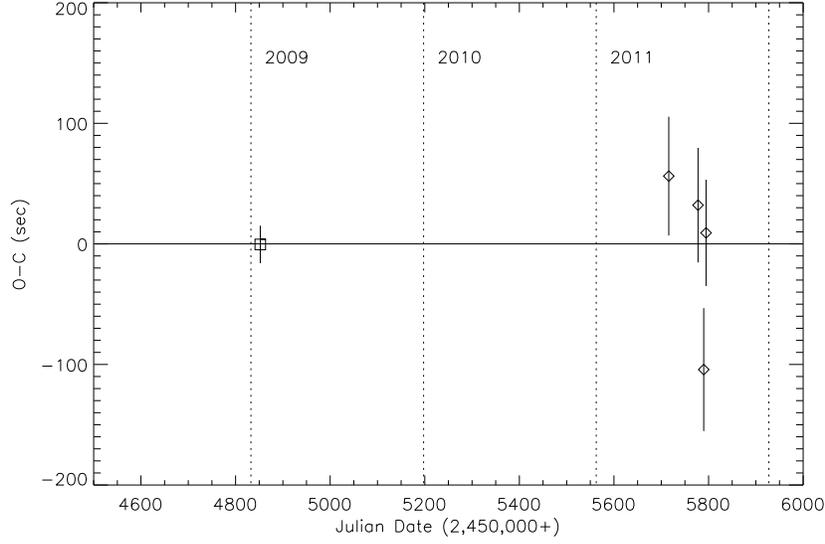}
	\caption{Timing residuals for HAT-P-23. Diamonds represent our observations and the square is the original ephemeris from \citep{Bakos10}. The weighed least-squares fit yields a period of $1.2128868 \pm 0.0000004$ days and epoch of $2,454,852.26542 \pm 0.00018$ (JDB\_TDB).}
	\label{fig:OC-residuals}
\end{figure}

We derived an improved orbital period for the HAT-P-23 system by performing a least-squares linear fit to our data and the discovery ephemeris from \citet{Bakos10}, weighing the individual mid-transit times by their uncertainties. We have converted the \citet{Bakos10} Barycentric Julian Date based on Coordinated Universal Time (BJD\_UTC) to the improved Dynamical Time-based system (BJD\_TDB) as suggested, for example, in \citet{Eastman10}. The difference between UTC-based and TDB-based timings is a systematic offset which depends on recent additions of leap seconds to UTC. To convert BJD\_UTC to BJD\_TDB 0.00077 days must be added for transits observed after 2009 January 1st (JD 2,454,832.5). Our analysis yields an ephemeris for the system with a period of $1.2128868 \pm 0.0000004$ days and an epoch of $2,454,852.26542 \pm 0.00018$ (BJD\_TDB). The timing residuals are shown in Fig.~\ref{fig:OC-residuals}. Our result improves the period reported in the discovery paper by \citet{Bakos10} and also reduces the uncertainty. No transit-timing variations (TTVs) from a single period, which would suggest the presence of other planets in the system, can be claimed at this time.

\section{Transit Modeling}
\label{sec:transit-modeling}

To estimate the essential transit parameters (the planet-to-star radius ratio: $R_{p}/R_{\star}$, the scaled semi-major axis: $a/R_{\star}$, and the orbital inclination of the system: $i$) from our combined ligh curve, we use the Transit Analysis Package (TAP) software package designed for IDL \citep{Gazak11}. It uses the now standard \citet{Mandel02} models for the transit light curve calculation and implements a Markov Chain Monte Carlo (MCMC) algorithm to find the best-fit parameters for the observed light curve. This analysis evaluates different combinations of parameters until it converges to an optimum solution. The package also implements the wavelet-based noise treatment algorithm developed by \citet{Carter09}. Such a scheme also allows estimating the uncertainty of the transit parameters. The Monte Carlo method implemented in TAP is a Metropolis-Hastings algorithm (see for example \citet{Ford06}) that finds the set of parameters that minimizes the value of $\chi^2$ in the analyzed light curve. 

The program starts with an initial estimation of the parameters and modifies a parameter of the set randomly, within reasonable limits of the parameter space. If the new set is a better estimation, the program accepts it. If not, the new set can be accepted or rejected depending on a condition that is evaluated randomly. Each set of parameters is defined as a link, and after execution, the algorithm returns a chain of all the sets of parameters that were evaluated. In order to estimate the best-fit parameters, TAP performs a Bayesian inference on the resulting chains, provided that a state of convergence has been reached in each chain. All the chains are added together  and, for each parameter, TAP calculates the median, and the 15.9 and 84.1 percentile levels. The median is reported as the estimation of the parameter and the mentioned percentile levels are used to estimate the $1\sigma$ value.

The TAP software, however, does not perform a joint fit of the photometry data with the radial velocity data. This is a limitation in this case since the value of $a/R_{\star}$ also depends on the radial velocity parameters when the orbit is non-circular. The intrinsic geometrical parameter that comes out of the transits alone is the duration of the transit, equivalent to $\zeta/R_{\star}$ (reciprocal of the half duration of the transit) in \citet{Bakos10}. The relation between $a/R_{\star}$ is expressed by equation (1) in \citet{Bakos10}, and from this it is evident that there is a subtle dependency of $a/R_{\star}$ on the eccentricity and orientation of the orbit. Therefore, one cannot derive $a/R_{*}$ without knowing these orbital parameters. The dependency vanishes for circular orbits, but increases with eccentricity. This is the reason we adopted eccentricity $e$ and the longitude of periastron $\omega$ values reported in \citet{Bakos10}. The orbital period, also an important parameter for determinig $a/R_{\star}$, can be determined from radial velocity curves as well as by successive observations of transits, as explained in \S~\ref{subsec:ephemeris}.

To start de MCMC algorithm, we set the orbital parameters of the system as reported by \citet{Bakos10} $(e = 0.106 \pm 0.044, \omega = 118\arcdeg \pm 25\arcdeg)$ except for the period, where we used our result from the previous section, and quadratic limb-darkening coefficients for the star from \citet{Claret00}. For the case of HAT-P-23 ($T_{\mathrm{eff}} = 5905 \pm 50$, $\mathrm{[Fe/H]} = 0.15 \pm 0.04$, and $\log g = 4.33 \pm 0.06$ ---also from \citet{Bakos10}) the coefficients are $a = 0.324$ and $b = 0.339$ for our photometric bandpass. \citet{Southworth08} concluded that there is no significant difference in using either a single linear ($u$) or two quadratic ($a$ \& $b$) coefficients to describe the stellar limb-darkening in the analysis of high quality ground-based data. We ran 10 chains with $10^6$ samples in each chain. The results of our modeling are presented in Table~\ref{tab:model-parameters} along with the original parameters of \citet{Bakos10} for comparison.

To verify our results we have performed alternative model solutions. One alternative consisted of doing a simultaneous fit to the four individual light curves (instead of the combined one) searching for a global solution to our three main model parameters ($R_{p}/R_{\star}$, $a/R_{\star}$, and $i$). In this case we set the period and mid-transit times to the values obtained in \S~\ref{sec:light-curve-analysis} and used the light curves where the baseline had been corrected using the airmass function as outlined in \S~\ref{sec:obs-and-reduction}. Not surprisingly the results and uncertainties were very nearly the ones obtained originally given the random nature of the Monte Carlo fits. This, however, outlines the fact that our uncertainties may be optimistic in the sense that the uncertainties derived from calculating the mid-transit times, the period, and the airmass correction to the baseline are not carried over to the Monte Carlo solution. In order to obtain more realistic values to our final model uncertainties we need to solve simultaneously for these parameters as well. In this second analysis we again use the TAP software to search for a global solution to our three main model parameters, and now also let the software find the best values for the mid-transit times of the individual light curves ($T_{\mathrm{c}1}$, $T_{\mathrm{c}2}$, $T_{\mathrm{c}3}$, $T_{\mathrm{c}4}$), a baseline corrections for each one (though we are limited in the sense that TAP uses a linear fit and not one based on the airmass value of the observations), and the system period. After longer computational time the resulting model values are similar to the ones obtained in our original analysis, but the formal uncertainties have nearly doubled for the model parameters and tripled for the mid-transit times. These are included in Tables \ref{tab:mid-transit-times} and \ref{tab:model-parameters} for comparison.

The differences in the uncertainties between the two methods are expected. A “true” global fit would require all model parameters to be solved simultaneously with as much data as possible. This would also have to include additional parameters like the limb-darkening coefficients and the rest of the orbital elements of the system. Since our available data is limited we elected to fix as many of the parameters as possible using various methods before solving for the three parameters of interest. Our period determination, for example, is more accurate than the one that could be derived from our four light curves alone since we use a longer time baseline. By the same token, our mid-transit times are also optimized to the individual light curves and solved for using robust methods. Individual fitting of the light curves may also be warranted since simultaneously solving for a single period may not describe the system accurately in the presence of possible mid-transit timing variations due to the presence of another planet in the system. We also feel that our airmass-dependent light curve baseline correction is a stronger model than a simple linear fit. Our guiding principle was the optimization of the individual parameter fits by minimizing the resulting uncertainties. After constraining those parameters as best as possible using independent methods, then we proceeded to solve for the remaining ones.

\section{Discussion}
\label{sec:discussion}

The results presented in Table~\ref{tab:model-parameters} agree in general with those found initially by \citet{Bakos10}. In some instances, we managed to decrease the uncertainty levels. In particular, our modeled inclination ($i$) and scaled semi-major axis ($a/R_*$) for the system agrees with the results obtained by \citet{Bakos10} within the $1 \sigma$ level of the measurements. While our inclination uncertainty could be considered slightly larger than \citet{Bakos10}, our scaled semi-major axis uncertainty is reduced by at least a factor of two. We obtain a more central transit and a slightly larger distance of the planet from the star. In general, there is degeneracy in these two variables since a smaller impact parameter (an inclination closer to $90 \arcdeg$) results in a longer transit duration, which is also the result of a larger semi-major axis. This effect can also be mimicked by a planet further away from its star (thus going slower) and having a larger impact parameter (crossing less surface area of the star). Limb-darkening effects observed in light curves taken at different wavelengths could help us sort out which case better represents this system. This comes about because stellar limb-darkening in general decreases with the inverse of the wavelength, yielding “flatter” depths in the light curves and “steeper” ingress and egress profiles. Our results do not show the sense of this degeneracy which leads us to believe that they may be real, although observations of transits at other wavelengths, preferably in the mid-infrared (J-, H-, K-band) are still desirable to settle this question. The major discrepancy in our results is the scaled planet-to-star radius ratio ($Rp/R_*$) which is smaller in our case and with similar uncertainties as \citet{Bakos10}. This is evidenced by a less deep transit depth in the light curve. Given a stellar radius, our planet is $\sim 5.5\%$ smaller, but we must point out that our results would still overlap at the $3 \sigma$ level of uncertainty. This agrees with the expectation of \citet{Fortney08} which predicted a smaller planet size ($\sim 8.4\%$ smaller) than observed by \citet{Bakos10} from their theoretical models for a planet of this mass.

\section{Conclusion}
\label{sec:conclusion}

We have shown in the present work that several exoplanet transit light curves obtained with small telescopes can be successfully combined to produce a higher quality transit light curve that can then be modeled and compared with observations performed with telescopes of larger aperture. Our results in general compare favorably with those of \citet{Bakos10}, with instances of improved uncertainties. In particular, we observe a smaller planet size which agrees better with the expectations of \citet{Fortney08}, although it must be emphasized that there are still large uncertainties in planetary models and no firm conclusion can be obtained from the results presented here. However, we could not improve on the uncertainty regarding the inclination of the system, which would have helped answer the question regarding the projected angle between the orbital plane and the stellar equatorial plane outlined in \citet{Moutou11}. Furthermore, the individual light curve observations are valuable in the sense that their mid-transit times can be accurately determined and this in turn would help refine the orbital period of the system and/or determine possible transit timing variations that could hint at the presence of another exoplanet in the system. An accurate orbital period is also valuable because it can be compared with the timings of secondary eclipses of the system. Comparing these values gives a more accurate measure of the eccentricity and argument of perihelion of the system than the radial velocity method. HAT-P-23b in particular is an inflated hot Jupiter-type planet in a close-in eccentric orbit that would show relatively deep secondary eclipse light curves that may be observed with ground-based instruments, and thus an accurate period is desirable for this system. Further observations of transits and possible secondary eclipses at various wavelengths with either larger telescopes or combined smaller telescopes are desirable in order to improve on the model parameters and better characterize this important inflated hot Jupiter system.

The authors are grateful to an anonymous referee for the insightful comments on this work.

\begin{table}[!t]
  \centering
  \setlength{\tabnotewidth}{0.75\linewidth}
  \setlength{\tabcolsep}{2.5\tabcolsep}
  \tablecols{3}
  \caption{Mid-Transit Times for HAT-P-23} 
  \label{tab:mid-transit-times}
  \begin{tabular}{lcc}
    \toprule
    \multirow{3}{*}{UT Date} & BM3 & Global Fit \\
	& Mid-Transit Time & Mid-Transit Time \\
	& BJD\_TDB \tabnotemark{a} & BJD\_TDB \tabnotemark{a} \\
    \midrule
    3 June 2011 & $5715.84150 \pm 0.00057$ & $5715.84202^{+0.00139}_{-0.00140}$ \\
    4 August 2011 & $5777.69845 \pm 0.00055$ & $5777.69800^{+0.00143}_{-0.00153}$ \\
    16 August 2011 & $5789.82574 \pm 0.00051$ & $5789.82612^{+0.00144}_{-0.00140}$ \\
    21 August 2011 & $5794.67860 \pm 0.00051$ & $5794.67746^{+0.00156}_{-0.00165}$  \\
    \bottomrule
    \tabnotetext{a}{Barycentric Julian Date based on Dynamical Time (2,450,000+). To convert to BJD\_UTC (Coordinated Universal Time) subtract 0.00077 days.}
  \end{tabular}
\end{table} 

\begin{table}[!t]
    \centering
    \caption{Model System Parameters for HAT-P-23}
    \setlength{\tabnotewidth}{0.95\linewidth}
    \setlength{\tabcolsep}{2.2\tabcolsep}
    \tablecols{4}
    \label{tab:model-parameters}
    \begin{tabular}{lccc}
      \toprule
      \multirow{2}{*}{Parameter} & Combined & Global Fit & Bakos et~al.\@ (2011) \\
	& Light Curve & &  \\
      \midrule
      $i$ & $87.9 ^{+1.5}_{-2.2}$ & $87.2^{+1.9}_ {-2.0}$ & $85.1 \pm 1.5$ \\
      $a/R_*$ & $4.23 ^{+0.06}_{-0.12}$ \tabnotemark{a} & $4.32^{+0.11}_{-0.17}$ \tabnotemark{a} & $4.14 \pm 0.23$ \\
      $R_p/R_*$ & $0.1105 ^{+0.0015}_{-0.0013}$ & $0.1057^{+0.0031}_{-0.0026}$ & $0.1169 \pm 0.0012$ \\ 
      \bottomrule
      \tabnotetext{a}{This parameter was obtained using $e$ and $\omega$ values from \citet{Bakos10}.}
    \end{tabular}
\end{table}

\end{document}